\documentclass{ws-mpla}
\usepackage[super]{cite}
\usepackage{graphicx}
\usepackage{amsmath}
\usepackage{eurosym}
\usepackage{amssymb}
\usepackage{color}

\def\be{\begin{equation}}
\def\ee{\end{equation}}
\def\bea{\begin{eqnarray}}
\def\eea{\end{eqnarray}}

\begin{document}

\markboth{M. K. Mak, C. S. Leung, T. Harko}
{Dark energy effects on static Schr\"{o}dinger-Newton
system}

%%%%%%%%%%%%%%%%%%%%% Publisher's Area please ignore %%%%%%%%%%%%%%
\catchline{}{}{}{}{}
%%%%%%%%%%%%%%%%%%%%%%%%%%%%%%%%%%%%%%%%%%%%%%%%%%%%%%%%%%%%%%%%%%%

\title{{The effects of the dark energy on the static Schr\"{o}dinger-Newton
system - an Adomian Decomposition Method and Pad\'{e} approximants based
approach}
}

\author{Man Kwong Mak}

\address{Departamento de F\'{\i}sica, Facultad de Ciencias Naturales, Universidad de
Atacama, Copayapu 485, Copiap\'o, Chile,\\
{\it email: mankwongmak@gmail.com}}

\author{Chun Sing Leung}

\address{Department of Applied Mathematics, Hong Kong Polytechnic University, Hong
Kong, Hong Kong SAR, P. R. China,\\
{\it email: chun-sing-hkpu.leung@polyu.edu.hk}}

\author{Tiberiu Harko}

\address{Astronomical Observatory, 19 Ciresilor Street, 400487 Cluj-Napoca, Romania,\\
{\it email: tiberiu.harko@aira.astro.ro}
}

\address{Department of Physics, Babes-Bolyai University, Kogalniceanu Street,
Cluj-Napoca 400084, Romania,
}

\address{School of Physics, Sun Yat-Sen University, Guangzhou 510275, People's
Republic of China
}

\maketitle

\pub{Received (25/092020)}{Revised (01/12/2020)}

\begin{abstract}
The Schr\"{o}dinger-Newton system is a nonlinear system obtained by coupling
together the linear Schr\"{o}dinger equation of quantum mechanics with the
Poisson equation of Newtonian mechanics. In the present work we will
investigate the effects of a cosmological constant (dark energy or vacuum
fluctuation) on the Schr\"{o}dinger-Newton system, by modifying the Poisson
equation through the addition of a new term. The corresponding Schr\"{o}%
dinger-Newton-$\Lambda$ system cannot be solved exactly, and therefore for
its study one must resort to either numerical or semianalytical methods. In
order to obtain a semianalytical solution of the system we apply the Adomian
Decomposition Method, a very powerful method used for solving a large class
of nonlinear ordinary and partial differential equations. Moreover, the
Adomian series are transformed into rational functions by using the Pad\'{e}
approximants. The semianalytical approximation is compared with the full
numerical solution, and the effects of the dark energy on the structure of
the Newtonian quantum system are  investigated in detail.


\end{abstract}

\ccode{PACS Nos.: 04.50.Kd, 04.20.Cv, 04.20.Fy}

%\tableofcontents

\section{Introduction}

The search for quantum gravity is one of the major directions of research in
theoretical physics. There are many proposals for building a quantum theory
of gravity, but achieving this goal seems to be still far away. For recent
reviews of the present status of quantum gravity see \cite%
{rev1,rev2,rev3,rev4}. However, the difficulties of quantizing general
relativity, and of quantum field theory in curved geometries have led to the
suggestion that perhaps a satisfactory description of quantum gravity may be
obtained from \textit{the unification of quantum mechanics and Newtonian
gravity} \cite{Diosi1}. Hence, in this approach, the basic equations of
quantum gravity can be formulated as \cite{Diosi1}
\begin{equation}  \label{1}
i\hbar \frac{\partial \psi \left(\vec{r},t\right)}{\partial t}=-\frac{\hbar
^2}{2m} \Delta \psi \left(\vec{r},t\right)+m\Phi \left(\vec{r},t\right)\psi
\left(\vec{r},t\right),
\end{equation}
and
\begin{equation}  \label{2}
\Delta \Phi \left(\vec{r},t\right)=4\pi G \rho _m \left(\vec{r},t\right),
\end{equation}
respectively, where $\hbar$ is Planck's constant, $G$ is the gravitational constant, $m$
is the particle mass, $\psi \left(\vec{r},t\right)$ is the particle wave
function, $\Phi \left(\vec{r},t\right)$ is the gravitational potential,
satisfying the Poisson equation (\ref{2}), and $\rho _m\left(\vec{r},t\right)$ is the
mass density. As for the gravitational potential one must assume, in this
formulation, that it is a stochastic quantity, with moments given by $\left<\Phi \left(%
\vec{r},t\right)\right>=\Phi_{cl} \left(\vec{r},t\right)$, and $\Phi \left(%
\vec{r},t\right)\Phi \left(\vec{r}^{\prime },t^{\prime }\right)-\Phi \left(%
\vec{r}^{\prime },t^{\prime }\right)\Phi \left(\vec{r},t\right)=\left(\hbar
G/\left|\vec{r}-\vec{r}^{\prime }\right|\right)\delta \left(t-t^{\prime
}\right)$ \cite{Diosi1}, respectively. The averaged value of the gravitational potential
can be obtained as $\left<\Phi\right>\sim \sqrt {\hbar ^2G/mR^3}$,  which gives, by taking into account that in the Newtonian limit $\left<\Phi\right>=Gm/R$, the condition $m^3R\sim \hbar ^2/G$. Hence it
turns out that the particle behavior is genuinely quantum if the condition $%
m^3R<<\hbar ^2/G\approx 10^{-47} \;\mathrm{cm\;g^3}$ is satisfied.

By
assuming that the mass density can be represented as $\rho _m \left(\vec{r}%
,t\right)=m\left|\psi \left(\vec{r},t\right)\right|^2$, the system of
equations (\ref{1}) and (\ref{2}) becomes the so-called Schr\"{o}%
dinger-Newton (or Schr\"{o}dinger-Poisson) system, whose properties have been
intensively investigated \cite%
{SN1,SN2,SN3,SN4,SN5,SN6,SN7,SN8,SN9,SN10,SN11,SN12}. The Schr\"{o}%
dinger-Newton system can also be obtained immediately as the nonrelativistic
limit of the semi-classical theory of gravity, that is, the theory in which
the gravitational field is considered classical, while the matter part is
quantized, with the field equations given by \cite{FT}
\begin{equation}
R_{\mu \nu}-\frac{1}{2}g_{\mu \nu}R=\frac{8\pi G}{c^4}\left<\psi \left|\hat{T%
}_{\mu \nu}\right|\psi\right>,
\end{equation}
where $\hat{T}_{\mu \nu}$ is the quantum mechanical operator associated to the energy-momentum
tensor, with its expectation value computed by choosing some appropriate 
quantum state. The Schr\"{o}dinger-Newton system can be reduced to a single
differential-integral equation, given by \cite{SN4},
\begin{equation}\label{intdif}
i\hbar \frac{\partial \psi\left(\vec {r}, t\right)}{\partial t} = - \frac{%
\hbar^2}{2m} \nabla^{2}\psi \left(\vec{r}, t\right) - Gm^2 \int {\frac{
\left|\psi \left(\vec {r}\;^{\prime },t\right)\right|^2}{\left|\vec {r}-%
\vec {r}\;^{\prime }\right|} \psi \left(\vec{r},t\right)d{\vec {r}\;^{\prime
}}} .
\end{equation}

The  Schr\"{o}dinger-Newton system with both local and nonlocal nonlinearities was investigated numerically in \cite{Gomes}, by also including in the model the modifications of the gravitational force due to the non-minimal coupling between curvature and matter, and  by using the numerical solvers developed for studying light propagating in the S-N model.

Recently, an extension of the standard  Schr\"{o}dinger-Newton system was proposed and investigated
in \cite{Matt}, by including in the mathematical formalism the effects of
the dark energy, represented by a cosmological constant $\Lambda$. Presently,
it is assumed that dark energy drives the late-time acceleration of the
Universe, and plays a determining role in the late cosmological evolution
\cite{Am}; for alternative models of dark energy as modified gravity see
\cite{RHL}, and references therein. In \cite{Matt} the regime in which dark energy
dominates both canonical quantum diffusion as well as gravitational self-attraction
was investigated in detail by numerically solving Eq.~(\ref{intdif}). It was found that the dark energy domination regime occurs for
sufficiently delocalized objects with an arbitrary mass. Moreover, one must also note that a minimal
delocalization width of about 67 m was determined from the high precision numerical analysis.
The modifications of an initially spherical Gaussian wave packet induced
by the presence of a positive cosmological constant and of the gravitational field were also investigated. It turns
out that the order of magnitude of the radial distance separating the collapsing phase from
the expansionary one is consistent with the analytical estimates obtained for the classical
turnaround radius for a spherically symmetric massive object in the presence of dark
energy. However, the physical time required to detect experimentally  these modifications is very
large, and therefore they can be measured only in physical systems containing a high
effective cosmological constant (dark energy), or, alternatively, via their effects in a
stationary  Universe.

It is the goal of the present Letter to investigate the mathematical and
physical properties of the \textit{static} Schr\"{o}dinger-Newton system in
the presence of dark energy, modeled as a cosmological constant. We call the
corresponding mathematical and physical model as the Schr\"{o}dinger-Newton-$%
\Lambda$ (S-N-$\Lambda$) system, and it represents a natural generalization
of the standard Schr\"{o}dinger-Newton model of quantum gravity. In order to
gain a better understanding of the physical and mathematical properties of
the S-N-$\Lambda$ system we will also obtain some semianalytical solutions
of it, by using the Adomian Decomposition Method. The Adomian Decomposition Method is a powerful mathematical
technique introduced in \cite{new2, R1, R2,b2}, and which was applied for
obtaining solutions of a large class of nonlinear ordinary, stochastic, and partial differential equations, or of
integral equations \cite{b5,b6,a1,a11, a2,a3,a4,a5,a6,a7,a8,a9}, with
applications in various scientific fields.

In our present approach we
first reformulate the static Schr\"{o}dinger-Newton-$\Lambda$ equations as a system of
two integral equations, and we  solve them by expanding the nonlinear
terms by using the Adomian polynomials \cite{new2,R1,R2,b2}. This allows us to
obtain a series solution of the S-N-$\Lambda$ system. In order to avoid the possible 
oscillatory or singular behavior of the solution we will represent the Adomian series
with the help of their Pad\'{e} approximants. The semianalytical results are
compared with the full numerical solutions for a large range of values of
the effective cosmological constant. We find that the Adomian-Pad\'{e} type
solutions give a good description of the numerical results for the static Schr\"{o}dinger-Newton system, and thus they
can offer some new insights into the important problem of the quantization
of gravity for static spherically systems at the Newtonian level.

The present Letter is organized as follows. In Section~\ref{sect2} we
present the Schr\"{o}dinger-Newton-$\Lambda$ system, and obtain its integral
representation. The Adomian Decomposition Method, as well as the Pad\'{e}
approximation is also briefly introduced. We obtain the semianalytical solution of the Schr%
\"{o}dinger-Newton-$\Lambda$ system for arbitrary initial conditions in
Section~\ref{sect3}. The comparison of the Adomian Decomposition Method results with the  full numerical solution is
performed in Section~\ref{sect4}. We discuss and conclude our results in Section~\ref{sect5}.

\section{The static Schr\"{o}dinger-Newton-$\Lambda$ system, and its
integral representation}\label{sect2}

In the present Section we  introduce the Schr\"{o}dinger-Newton-$\Lambda$
system, and we  present its dimensionless form. Moreover, we will obtain
the integral equations representation of the system. We  also
briefly introduce the Adomian Decomposition Method for systems of ordinary nonlinear differential equations.

\subsection{The static, spherically symmetric Schr\"{o}dinger-Newton-$%
\Lambda $ system}

For a single particle system the static Schr\"{o}dinger-Newton-$\Lambda$
system takes the form \cite{Matt}
\begin{equation}  \label{1a}
-\frac{\hbar ^2}{2m}\Delta \psi \left(\vec{r}\right) +m\Phi\left(\vec{r}%
\right)\psi \left(\vec{r}\right) =E\psi \left(\vec{r}\right),
\end{equation}
\begin{equation}  \label{1b}
\Delta \Phi \left(\vec{r}\right)=4\pi Gm \left|\psi \left(\vec{r}%
\right)\right|^2-\Lambda c^2,
\end{equation}
where $E$ is the energy eigenvalue of the particle, while the
constant term $\Lambda$ models the effects of the dark energy, or, alternatively,  of the
vacuum fluctuations. In the case $\Lambda \equiv 0$, the system of equations
(\ref{1a}) and (\ref{1b}) reduces to the standard static Schr\"{o}dinger-Newton system, whose interesting properties have been investigated in
detail \cite{T1,T2,T3,T3a,T4, T5, T6, T7}. In the following we will assume,
without any loss of generality, that the wave function $\psi$ is real. In
order to obtain a simpler form of the S-N-$\Lambda$ system, we
introduce two new functions $S$ and $V$, defined as
$\psi \left(\vec{r}\right)=\left(\frac{\hbar ^2}{8\pi Gm^3}%
\right)^{1/2}S\left(\vec{r}\right)$, and $ E-m\Phi=\frac{\hbar ^2}{2m}V\left(\vec{%
r}\right)$, respectively \cite{T2,T3}\;. Both $S$ and $V$ have the physical units of 1/length$^2$. Then the Schr\"{o}dinger-Newton-$\Lambda$ system takes the form
\begin{equation}  \label{2a}
\Delta S =-SV,
\end{equation}
\begin{equation}
\Delta V=-S^2+\lambda,
\end{equation}%
\label{2b} where we have denoted
\begin{equation}
\lambda =\frac{2m^2c^2}{\hbar ^2}\Lambda=4.522\times 10^{-29}\times \left(\frac{m}{m_p}\right)^2\times \left(\frac{\Lambda}{10^{-56}\;{\rm cm^{-2}}}\right)\; {\rm cm^{-4}},
\end{equation}
where $m_p$ denotes the mass of the proton.

Eq.~(\ref{1a}) can be derived from the variational principle \cite{T4},
\be
H[\psi \left(\vec{r}\right)]=\frac{\hbar ^2}{2m}\int{\left[\left|\nabla \psi \left(\vec{r}\right)\right|^2+\frac{m}{4}\Phi\left(\vec{r}\right)\left|\psi \left(\vec{r}\right)\right|^2\right]d^3\vec{r}},
\ee
with the wave function satisfying the normalization condition $\int{\left|\psi\left(\vec{r}\right)\right|^2d^3\vec{r}}=1$. The Poisson equation (\ref{1b}) can be solved by using a Green function techniques to give \cite{Matt}
\be
\Phi\left(\vec{r}\right)=-Gm\int{\frac{\left|\psi \left(\vec{r}\right)\right|^2}{\left|\vec{r}-\vec{r}\;'\right|}d^3\vec{r}\;'}+\frac{\Lambda c^2}{4\pi}\int{\frac{1}{\left|\vec{r}-\vec{r}\;'\right|}d^3\vec{r}\;'}.
\ee
Hence, the Schr\"{o}dinger-Newton-$\Lambda$ system can be obtained as an extremum of the functional
\bea\label{act}
H\left[\psi \left(\vec{r}\right)\right]&=&\frac{\hbar ^2}{2m}\int{\left|\nabla \psi \left(\vec{r}\right)\right|^2d^3\vec{r}}-Gm^2\int{\int{\frac{\left|\psi \left(\vec{r}\right)\right|^2\left|\psi \left(\vec{r}\;'\right)\right|^2}{\left|\vec{r}-\vec{r}\;'\right|}d^3\vec{r}d^3\vec{r}\;'}}+\nonumber\\
&&\frac{m c^2}{4\pi}\Lambda\int{\int{\frac{\left|\psi \left(\vec{r}\right)\right|^2}{\left|\vec{r}-\vec{r}\;'\right|}d^3\vec{r}d^3\vec{r}\;'}}.
\eea

In the following we will limit our investigations to the spherically
symmetric case, with $S=S(r)$ and $V=V(r)$, respectively, where $r$ is the
radius vector. Moreover, for the wave function normalization as well as for
the energy eigenvalues we will adopt the same prescription as in the case of
the simple Schr\"{o}dinger-Newton system, namely, $\int_0^{\infty}{r^2S^2dr}=\frac{2Gm^3}{\hbar ^2}$, and $E=\frac{\hbar ^2}{2m}\lim _{r\rightarrow \infty}V(r)$, respectively \cite{T2,T3}.

 In the case of spherical symmetry the Schr\"{o}dinger-Newton-$%
\Lambda$ system can be written as
\begin{equation}  \label{3a}
\frac{1}{r}\frac{d^2}{dr^2}\left[rS(r)\right]=S^{\prime \prime }(r)+\frac{2}{%
r}S^{\prime }(r)=-S(r)V(r),
\end{equation}
\begin{equation}  \label{3b}
\frac{1}{r}\frac{d^2}{dr^2}\left[rV(r)\right]=V^{\prime \prime}(r)+\frac{2}{r%
}V^{\prime }(r)=-S^2(r)+\lambda.
\end{equation}
The system of equations (\ref{3a}) and (\ref{3b}) is invariant with respect to the transformations $S\rightarrow S/a^2$, $V\rightarrow V/a^2$, $r\rightarrow ar$, and $\lambda \rightarrow \lambda /a^4$, respectively, where $a={\rm constant}$. In the following we will consider the system (\ref{3a})-(\ref{3b}) with the
initial conditions $S(0)\neq 0$ and $V(0)\neq 0$, implying that the functions $%
S(r)$ and $V(r)$ are smooth and finite at the origin $r=0$, and $S^{\prime
}(0)=0$ and $V^{\prime }(0)=0$, respectively, that is, with vanishing
derivatives of $S$ and $V$ at the origin.

\subsection{Integral equation formulation of the Schr\"{o}dinger-Newton-$%
\Lambda$ system}

By integrating once Eqs.~(\ref{3a}) and (\ref{3b}) we obtain
\begin{equation}
\frac{d}{dr}\left[rS(r)\right]=S_0-\int_0^r{x^{\prime }S\left(x^{\prime
}\right)V\left(x^{\prime }\right)dx^{\prime }},
\end{equation}
and
\begin{equation}
\frac{d}{dr}\left[rV(r)\right]=V_0+\frac{\lambda r^2}{2}-\int_0^r{x'S^2\left(x^{\prime}\right)dx^{\prime }},
\end{equation}
respectively. A second integration gives
\begin{equation}
rS(r)=S_0r-\int_0^r{\int_0^{x^{\prime \prime }}{x^{\prime }S\left(x^{\prime
}\right)V\left(x^{\prime }\right)dx^{\prime }}dx^{\prime \prime }},
\end{equation}
\begin{equation}
rV(r)=V_0r+\frac{\lambda r^3}{6}-\int_0^r{\int_0^{x''}{x'S^2\left(x'\right)dx'}dx''}.
\end{equation}
By using the Cauchy formula for repeated integration, $\int_a^x\int_a^{x_1}...\int_a^{x_{n-1}}f\left(x_n\right)dx_n...dx_2dx_1=\left[1/(n-1)!\right]\int_a^x{(x-t)^{n-1}f(t)dt}$,  we finally obtain the
integral equation formulation of the Schr\"{o}dinger-Newton-$\Lambda$ system
as
\begin{equation}  \label{10}
S(r)=S(0)-\int_0^r{x\left(1-\frac{x}{r}\right)S(x)V(x)dx},
\end{equation}
and
\begin{equation}  \label{11}
V(r)=V(0)+\frac{\lambda r^2}{6}-\int_0^r{x\left(1-\frac{x}{r}\right)S^2(x)dx}%
,
\end{equation}
respectively.  By taking the derivative of Eq.~(\ref{11}) with respect to $r$ we obtain
\be
V'(r)=\frac{\lambda r}{3}-\int_0^r{\frac{x^2S^2(x)}{r^2}dx}=\frac{\lambda r}{3}-\phi(r),
\ee
where $\phi(r)=\int_0^r{x^2S^2(x)dx/r^2}$. According to a standard result in calculus, if $f:I\rightarrow R$ is a continuous and positive function on $I$, then $\int_I{f(x)dx}>0$. Since obviously $x^2S^2(x)/r^2>0$, it follows that $\phi (r)>0, \forall r>0$. Hence for $\lambda =0$,  $V^{\prime }(r)<0$, and, therefore, in the absence of dark energy $V$ must be a monotonically decreasing function of $r$. However, there is a drastic change in the behavior of $V(r)$ in the presence of the cosmological constant $\lambda$. If $\lambda$ satisfies the condition $\lambda <3\phi (r)/r, \forall r\geq 0$, then, similarly to the standard Schr\"{o}dinger-Newton case, $V'(r)<0, \forall r>0$, $V(r)$ is a monotonically decreasing function of the radial coordinate, and, if $V$ diverges at infinity, then $\lim_{r\rightarrow \infty}V(r)=-\infty$. On the other hand, if $\lambda >3\phi (r)/r, \forall r\geq 0$, $V'(r)>0,\forall r>0$, and $V(r)$ is a monotonically increasing function of $r$. Consequently, if $V$ is singular at infinity, then $\lim_{r\rightarrow \infty}V(r)=+\infty$. Generally, the rescaled gravitational potential $V(r)$ satisfies in the presence of the cosmological constant  the condition
\be
\frac{d}{dr}\left[V(r)-V(0)-\frac{\lambda r^2}{6}\right]<0,
\ee
which generalizes the condition $V'(r)<0$ valid for the Schr\"{o}dinger-Newton system.

\subsection{The Adomian Decomposition Method}

We  illustrate now the Adomian Decomposition Method for the case of a
nonlinear second order ordinary differential equation, written in 
Adomian's operator-theoretic notation as \cite{new2, R1, R2,b2}
\begin{equation}  \label{12}
Lu(x)+Ru(x)+Nu(x)=g(x),
\end{equation}
where $g(x)$ is the system input, $u(x)$ is the system output, $L$ is the
highest order differential operator, given, in our case, by $L(.)=\frac{d^2}{%
dx^2}(.)$, $R$ is the linear operator, while $N$ is the nonlinear operator,
assumed to be analytic. In order to solve the initial value problem
associated to Eq.~(\ref{12}), we adopt for the inverse linear operator $%
L^{-1} $ the two-fold definite integral $L^{-1}(.)=\int_a^x \int_a^x(.)dxdx$%
, where $a$ is the initial point. By applying the inverse of the operator $L$
to both sides of Eq.~(\ref{12}) we obtain the formal solution of the nonlinear differential
equation as \cite{new2, R1, R2,b2}
\begin{equation}
u=U+L^{-1}g-L^{-1}\left(Ru+Nu\right),
\end{equation}
where the first term $U$ in the above relation contains the initial conditions, and identically satisfies the
equation $LU=0$.

The basic idea of the Adomian Decomposition Method is to
represent $u(x)$ by the Adomian Decomposition series, $u(x)=\sum
_{n=0}^{\infty}{u_n(x)}$, while the nonlinear term $Nu$ is represented in
terms of the Adomian polynomials $A_n(x)$, given by the formal expression, $Nu(x)=\sum _{n=0}^{\infty}{A_n(x)}$. For a nonlinearity
of the form $Nu=f(u)$ the $A_n$'s are defined according to \cite{new2, R1, R2,b2}
\begin{equation}
A_n=A_n\left(u_0,u_1,...,u_n\right)=\frac{1}{n!}\left.\frac{d^n}{d\lambda ^n}%
f\left(\sum _{k=0}^n{\lambda ^ku_k(x)}\right)\right|_{\lambda =0}.
\end{equation}

Substituting the Adomian expansions into Eq.~(\ref{12}) we obtain the
following recursion scheme for the solution components,
\begin{equation}
u_0=U+L^{-1}g,
\end{equation}
\begin{equation}
u_{n+1}=-L^{-1}\left(Ru_n+A_n\right), n\geq 0.
\end{equation}
Hence the $n+1$-term approximation of the solution is $u_{n+1}=\sum
_{k=0}^nu_k$. In order to obtain a better approximation of the solution we
will use the Pad\'{e} approximants \cite{b5} of the Adomian series, which
transform the polynomial approximations into a rational function that allows
us to obtain more information about $u(x)$. The Pad\'{e} approximants will
converge on the entire real axis if $u(x)$ is free of singularities \cite{b5}%
.

\section{Series solution of the Schr\"{o}dinger-Newton-$\Lambda$ system via
the Adomian Decomposition Method}\label{sect3}

In the present Section we will consider a semianalytical approach to the Schr%
\"{o}dinger-Newton-$\Lambda$ system, by using the Adomian Decomposition Method and the
Pad\'{e} approximants.

%\subsection{The Adomian Decomposition Method for the S-N-$\Lambda$ system}

In order to apply these mathematical methods we will consider the
equivalent mathematical formulations of the Schr\"{o}dinger-Newton-$\Lambda $
system as a system of integral equations. In the following we will look for
a series solution of the system (\ref{10}) and (\ref{11}), by assuming that $%
S(r)=\sum_{n=0}^{\infty }{S_{n}(r)}$, and $V(r)=\sum_{n=0}^{\infty }{V_{n}(r)%
}$, respectively. As for the nonlinear terms $S(x)V(x)$ and $S^{2}(x)$, we
will decompose them in terms of the Adomian polynomials according to
\begin{equation}
S(x)V(x)=\sum_{n=0}^{\infty }{A_{n}(x)},S^{2}(x)=\sum_{n=0}^{\infty }{%
B_{n}(x)}.
\end{equation}%
Substituting the above decompositions into Eqs.~(\ref{10}) and (\ref{11}) we
obtain
\begin{equation}
\sum_{n=0}^{\infty }{S_{n}(r)}=S(0)-\sum_{n=0}^{\infty }{\int_{0}^{r}{%
x\left( 1-\frac{x}{r}\right) A_{n}(x)dx}},
\end{equation}%
\begin{equation}
\sum_{n=0}^{\infty }{V_{n}(r)}=V(0)+\frac{\lambda r^{2}}{6}%
-\sum_{n=0}^{\infty }{\int_{0}^{r}{x\left( 1-\frac{x}{r}\right) B_{n}(x)dx}}.
\end{equation}%
Hence we obtain the following recursive relations for the determination of
the solution of the Schr\"{o}dinger-Newton-$\Lambda$ system,
\begin{equation}
S_{0}=S(0),S_{n+1}=-\int_{0}^{r}{x\left( 1-\frac{x}{r}\right) A_{n}(x)dx},
\end{equation}%
\begin{equation}
V_{0}=V(0)+\frac{\lambda r^{2}}{6},V_{n+1}=-\int_{0}^{r}{x\left( 1-\frac{x}{r%
}\right) B_{n}(x)dx}.
\end{equation}%
As for the Adomian polynomials, they are given by $A_{0}=S(0)V_{0}$, $%
A_{1}=S(0)V_{1}+S_{1}V_{0}$, $A_{2}=S(0)V_{2}+S_{1}V_{1}+S_{2}V_{0}$ etc.,
and $B_{0}=S^{2}(0)$, $B_{1}=2S(0)S_{1}$, $B_{2}=2S(0)S_{2}+S_{1}^{2}$, $%
B_{3}=2S(0)S_{3}+2S_{1}S_{2}$, respectively. Hence we obtain the first five
successive terms in the Adomian series expansion of the Schr\"{o}%
dinger-Newton-$\Lambda $ model as
\begin{equation}
S_{1}(r)=-\frac{1}{120}r^{2}S(0)\left[ \lambda r^{2}+20V(0)\right],
\ee
\be
V_{1}(r)=-\frac{1}{6}r^{2}S^{2}(0),
\end{equation}%
\begin{equation}
S_{2}(r)=\frac{r^{4}S(0)\left\{ \lambda r^{2}V(0)+42\left[ S^{2}(0)+V^{2}(0)%
\right] \right\} }{5040},
\ee
\be
V_{2}(r)=\frac{r^{4}S^{2}(0)\left[ \lambda
r^{2}+42V(0)\right] }{2520},
\end{equation}%
\begin{equation}
S_{3}(r)=-\frac{r^{6}S(0)\left\{ \lambda r^{2}\left[ 9S^{2}(0)+V^{2}(0)%
\right] +456S^{2}(0)V(0)+72V^{3}(0)\right\} }{362880},
\end{equation}%
\begin{equation}
V_{3}(r)=-\frac{r^{6}S^{2}(0)\left\{ 63\lambda ^{2}r^{4}+4400\lambda
r^{2}V(0)+13200\left[ 3S^{2}(0)+8V^{2}(0)\right] \right\} }{99792000},
\end{equation}%
\bea
S_{4}(r)&=&\frac{r^{8}S(0)}{15567552000}\Bigg\{ 393\lambda ^{2}r^{4}S^{2}(0)+78\lambda
r^{2}V(0)\left[ 557S^{2}(0)+5V^{2}(0)\right] +\nonumber\\
&&14300\left[
27S^{4}(0)+98S^{2}(0)V^{2}(0)+3V^{4}(0)\right] \Bigg\} ,
\eea
\bea
V_{4}(r)&=&\frac{r^{8}S^{2}(0) }{7783776000}\Bigg\{ 165\lambda ^{2}r^{4}V(0)+78\lambda r^{2}%
\left[ 171S^{2}(0)+191V^{2}(0)\right] +\nonumber\\
&&114400V(0)\left[ 5S^{2}(0)+3V^{2}(0)%
\right] \Bigg\}.
\eea
%\begin{eqnarray}
%S_{5}(r) &=&\frac{r^{10}S(0)}{6224529991680000}\Bigg\{-120393\lambda
%^{3}r^{6}S^{2}(0)-17719984\lambda ^{2}r^{4}S^{2}(0)V(0)-66640\lambda r^{2}%
%\Bigg[7461S^{4}(0)+  \notag \\
%&&16732S^{2}(0)V^{2}(0)+15V^{4}(0)\Bigg]-10395840V(0)\Bigg[%
%2791S^{4}(0)+2626S^{2}(0)V^{2}(0)+15V^{4}(0)\Bigg]\Bigg\},
%\end{eqnarray}%
%\begin{eqnarray}
%V_{5}(r) &=&\frac{r^{10}S^{2}(0)}{22884301440000}\Bigg\{\lambda ^{2}r^{4}%
%\Bigg[-\left( 50547S^{2}(0)+9295V^{2}(0)\right) \Bigg]-980\lambda r^{2}V(0)%
%\Bigg[5703S^{2}(0)+1135V^{2}(0)\Bigg]-  \notag \\
%&&305760\Bigg[S^{2}(0)+6V^{2}(0)\Bigg]\Bigg[81S^{2}(0)+16V^{2}(0)\Bigg]%
%\Bigg\}.
%\end{eqnarray}%
The next terms of the Adomian series expansion can be easily calculated. The
Pad\'{e} approximants of order $[3/4]$ of the Adomian series truncated to
the first six terms are given by
\bea
S(r)\left[ \frac{3}{4}\right] &=&\frac{1}{P(r)}\Bigg\{60r^{2}S(0)V(0)\left[ -39\lambda
+23S^2(0)-31V^2(0)\right] -\nonumber\\
&&2520S(0)\left[ -3\lambda
+3S^2(0)-7V^2(0)\right]\Bigg\} ,
\eea
and
\begin{equation}
V(r)\left[ \frac{3}{4}\right] =\frac{X(r)}{Y(r) },
\end{equation}
respectively, where
\bea
P(r)&=&r^{4}\Bigg\{ 63\lambda ^{2}+63S^4(0)-2S^2(0)\left[ 63\lambda
+32V(0)^{2}\right] +33V^4(0)-96\lambda V^2(0)\Bigg\} +\nonumber\\
&&120r^{2}V(0)\left[
-9\lambda +S^2(0)+9V^2(0)\right] -2520\left[ -3\lambda
+3S^2(0)-7V^2(0)\right],
\eea
\bea
X(r)&=&30\Bigg\{r^{2}\left[
35S^{4}(0)-39S^{2}(0)V^{2}(0)+8V^{4}(0)-3\lambda V^{2}(0)\right] +\nonumber\\
&&42V(0)%
\left[ 3V^{2}(0)-5S^{2}(0)\right] \Bigg\},
\eea
and
\bea
Y(r)&=&15r^{4}S^{4}(0)+30V(0)\left\{
2V(0)\left[ 4r^{2}V(0)+63\right] -3\lambda r^{2}\right\} -\nonumber\\
&&S^{2}(0)\left\{
r^{4}\left[ 15\lambda +23V^{2}(0)\right] +540r^{2}V(0)+6300\right\},
\eea
respectively.

\section{The numerical analysis of the S-N-$\Lambda$ system}\label{sect4}

 In the present Section we also perform a numerical analysis of the Schr\"{o}dinger-Newton-$\Lambda$ system, and compare the numerical results with the semianalytical approximations obtained via the Adomian Decomposition Method.

 \subsection{Full numerical results}

\begin{figure*}[ht]
  \centering
  \includegraphics[width=6cm]{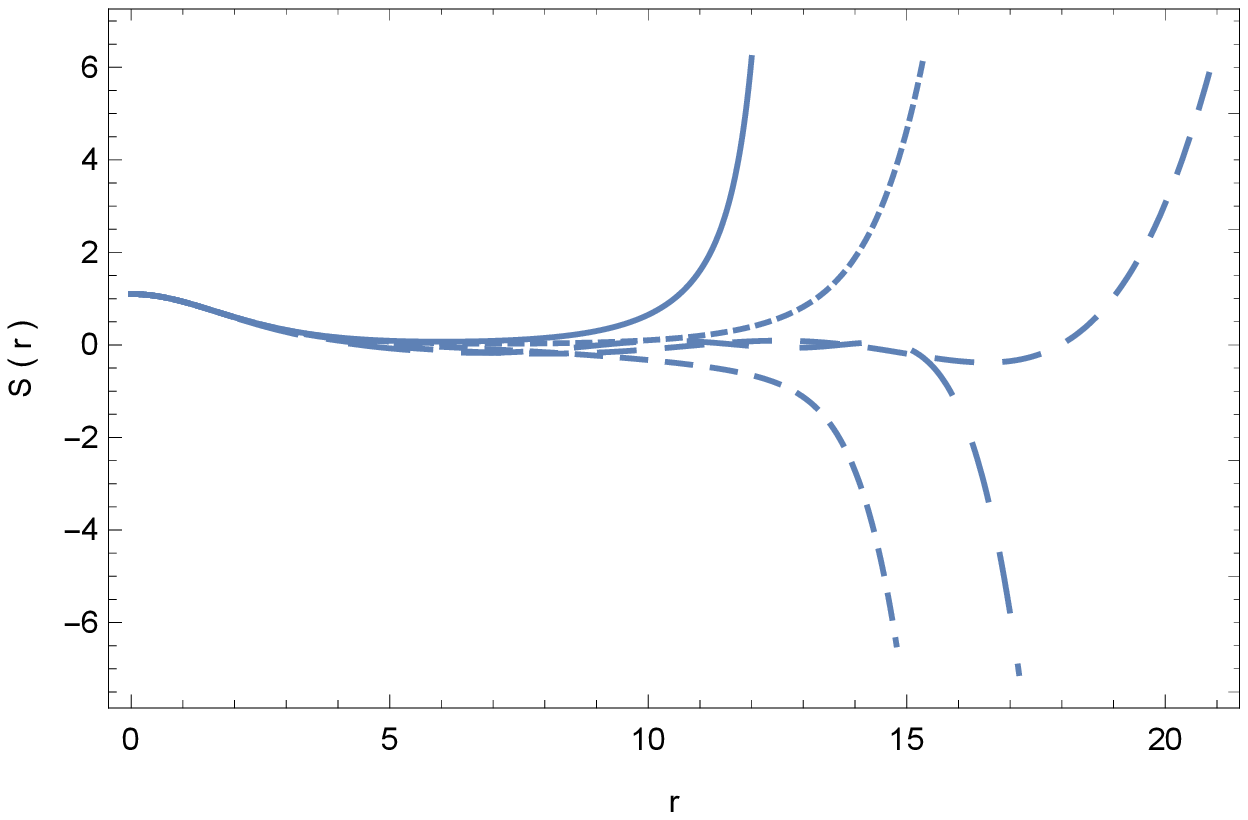}
  \includegraphics[width=6cm]{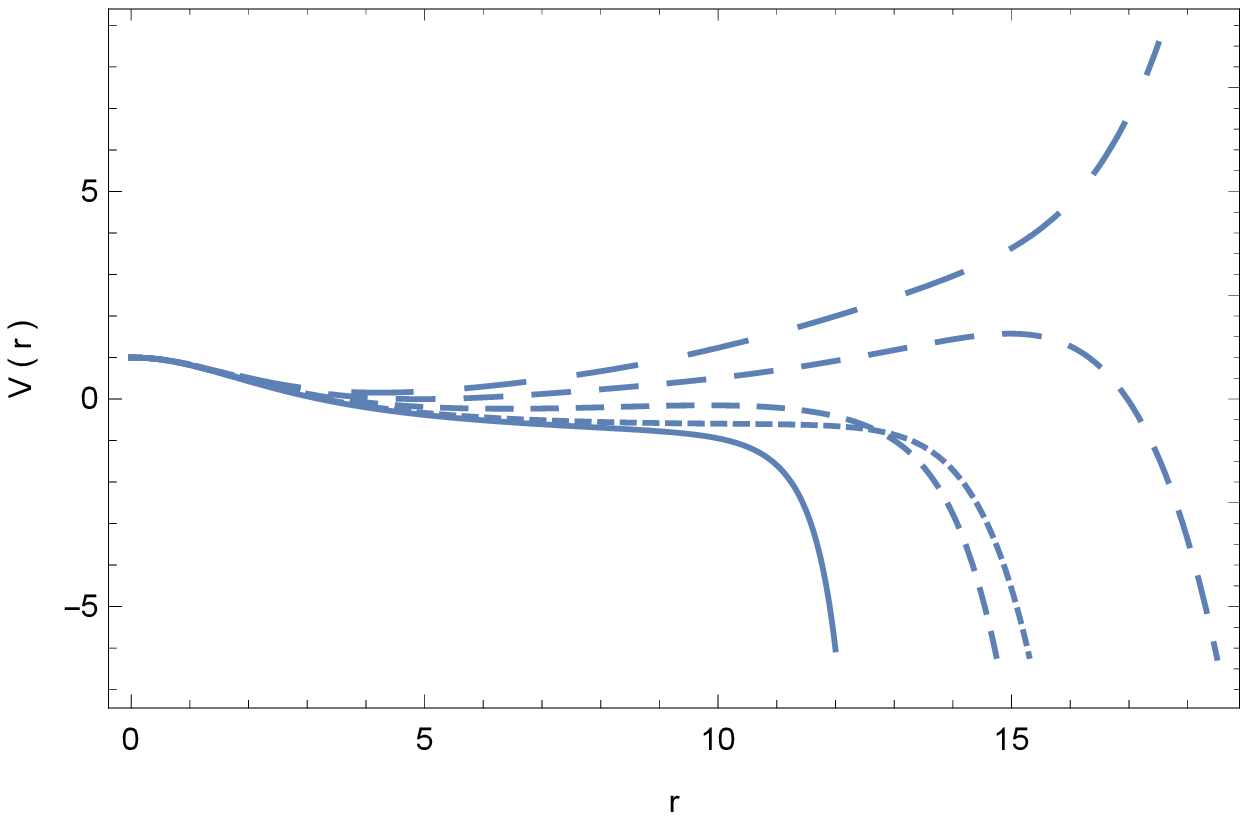}
  \caption{The behavior of $S(r)$ (left panel) and $V(r)$ (right panel) as a function of $r$ (in arbitrary units) for $S(0)=1.10$ and $V(0)=1.0$, and for different values of the cosmological constant: $\lambda =0$ (solid curve), $\lambda =0.01$ (dotted curve), $\lambda =0.04$ (short dashed curve), $\lambda =0.08$ (dashed curve), and $\lambda =0.12$ (long dashed curve).}\label{fig1}
\end{figure*}

 We will present first some full numerical results that indicate the effects of the cosmological constant on the behavior of the solutions of the S-N-$\Lambda$ system.  As it is already known from the numerical study of the Schr\"{o}dinger-Newton system, it admits solutions for which $\left(S, V\right)\rightarrow \left(\pm \infty, \pm \infty\right)$. For example, for the initial conditions $S(0)=1.10$ and $V(0)=1$, the system blows up at $r\approx 10$ so that  $S(r)\rightarrow +\infty$, and $V(r)\rightarrow -\infty$. However, with the inclusion of the cosmological constant in the model, both the quantitative and qualitative behavior of the model changes drastically, as shown in Fig.~\ref{fig1}.

The first effect of the cosmological constant is the significant modification of the position of the blow-up point, which significantly increases with the increase of $\lambda$. More importantly, a set of different blow-up solutions with $S(r)\rightarrow -\infty$ and $V(r)\rightarrow +\infty$ do appear. The position of the first zeros of $S$ and $V$ are also displaced, and it increases with increasing $\lambda$.

\begin{figure*}[ht]
  \centering
  \includegraphics[width=6cm]{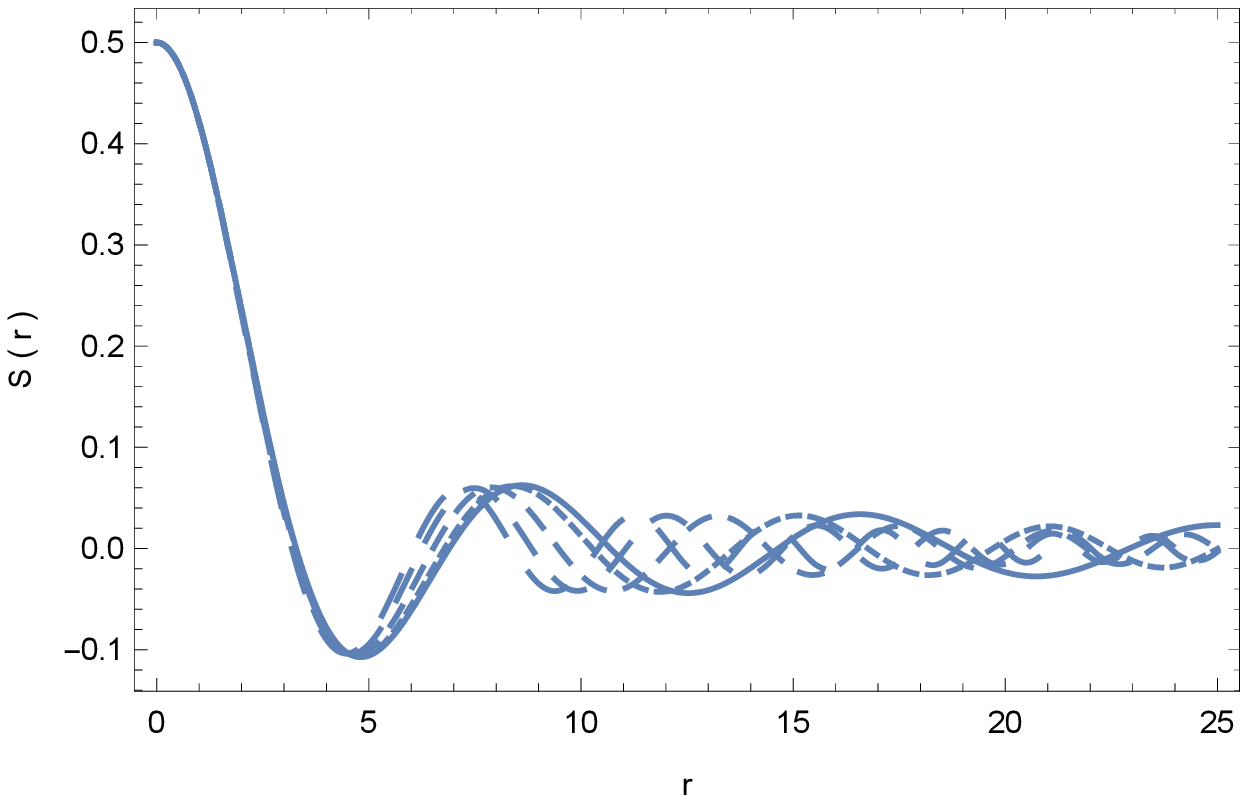}
  \includegraphics[width=6cm]{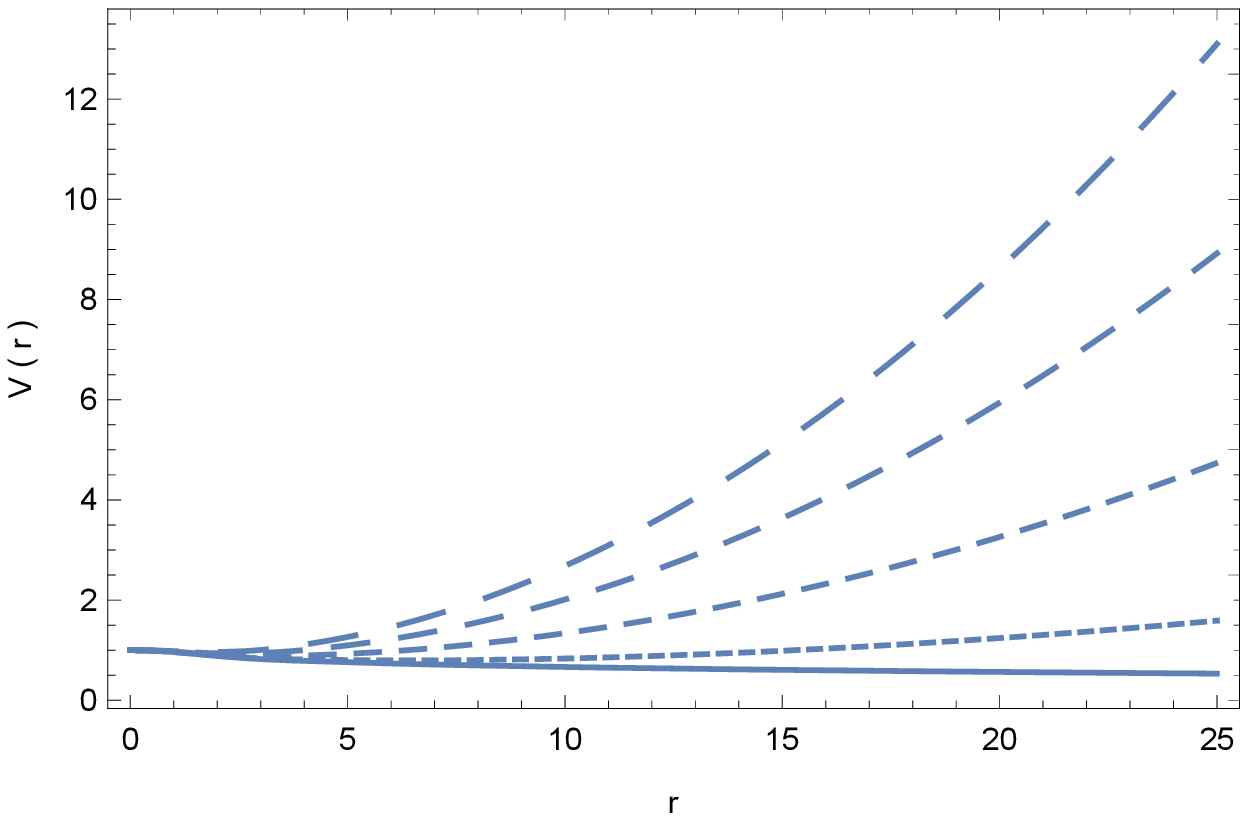}
  \caption{The behavior of $S(r)$ (left panel) and $V(r)$ (right panel) as a function of $r$ (in arbitrary units) for $S(0)=0.5$ and $V(0)=1.0$, and for different values of the cosmological constant: $\lambda =0$ (solid curve), $\lambda =0.01$ (dotted curve), $\lambda =0.04$ (short dashed curve), $\lambda =0.08$ (dashed curve), and $\lambda =0.12$ (long dashed curve).}\label{fig2}
\end{figure*}

 The variations of $S$ and $V$ for $S(0)=0.50$ and $V(0)=1.0$ are represented in Fig.~\ref{fig2}.

The presence of a cosmological constant has a significant impact on the behavior of the wave function, and on the gravitational potential energy. While the general oscillatory behavior of the wave function for large $r$ is not affected, the position of the zeros of $S$ depend  on the value of the cosmological constant. The positions of the maximums and minimums of $S$ are also displaced as compared to the $\lambda =0$ case. The effect of the dark energy on the effective gravitational potential is very significant. While for the simple S-N system $V$ is a monotonically decreasing function of $r$, tending to zero at infinity,  for the adopted values of the cosmological constant,  $V$ becomes a monotonically increasing function, diverging at infinity.

\begin{figure*}[ht]
  \centering
  \includegraphics[width=6cm]{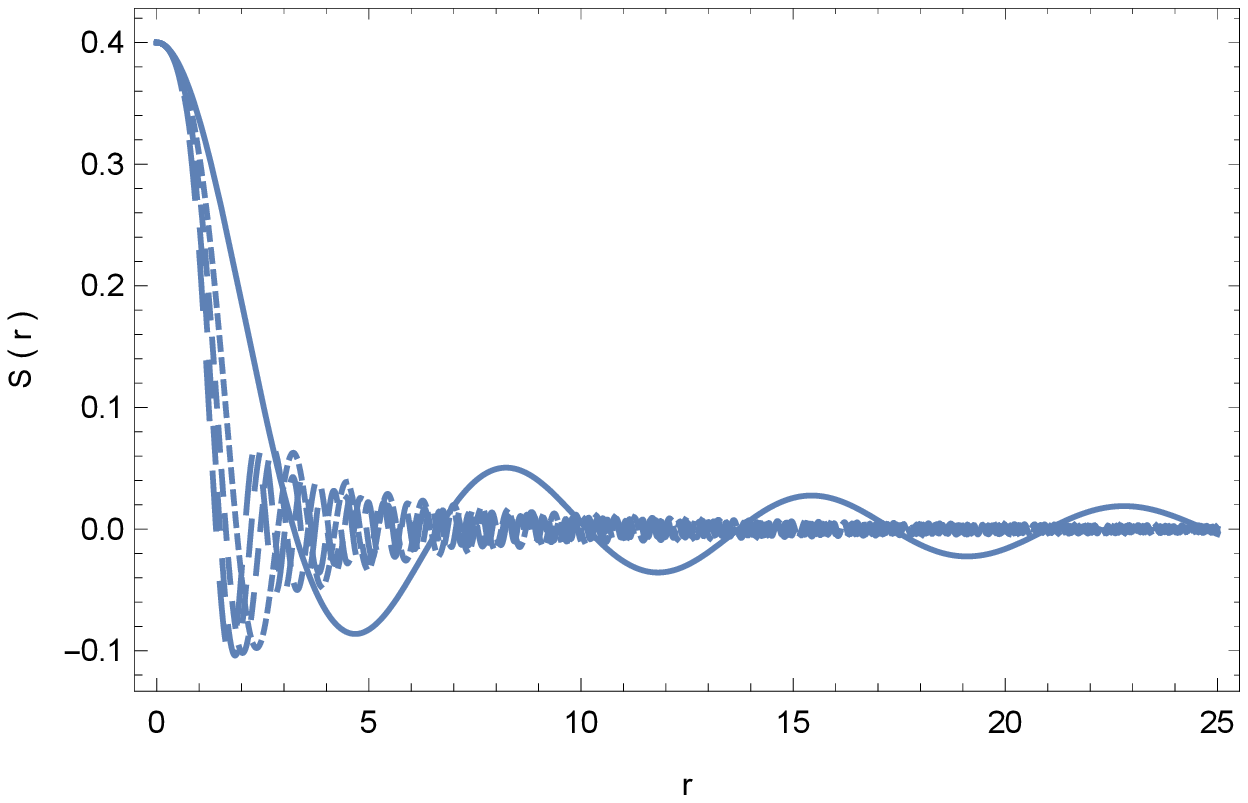}
  \includegraphics[width=6cm]{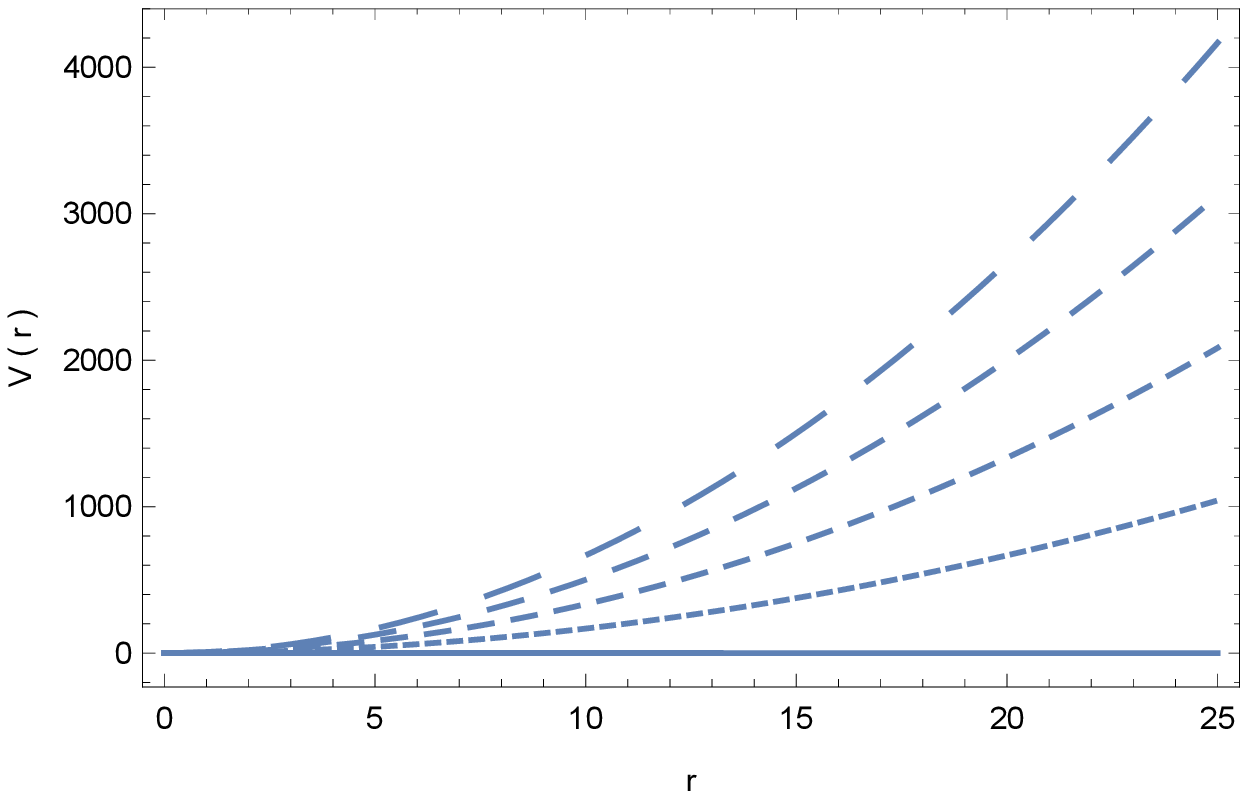}
  \caption{The behavior of $S(r)$ (left panel) and $V(r)$ (right panel) as a function of $r$ (in arbitrary units) for $S(0)=0.4$ and $V(0)=1.0$, and for different values of the dark energy: $\lambda =0$ (solid curve), $\lambda =10$ (dotted curve), $\lambda =20$ (short dashed curve), $\lambda =30$ (dashed curve), and $\lambda =40$ (long dashed curve).}\label{fig2a}
\end{figure*}

An interesting physical regime corresponds to the dark energy domination limit, corresponding to large values of $\lambda$. The comparison between the behavior of the wave function and gravitational potential of the N-S model with $\lambda =0$ and the dark energy dominated quantum system is presented for $S(0)=0.40$, $V(0)=1$, and for different large values of $\lambda$, in Fig.~\ref{fig2a}.

As one can see from the Figures, the presence of a large dark energy induces both  qualitative and quantitative differences as compared to the behavior of the standard S-N system. The oscillating behavior of the wave function is strongly modified, and in the presence of $\lambda$ the transition to zero takes place through a large number of oscillations. While for the S-N system $V$ is a slowly decreasing function of $r$, in the S-N-$\Lambda$ model $V(r)$ is a rapidly increasing function, taking numerical values three orders of magnitude higher than in the $\lambda =0$ case.

\begin{figure*}[ht]
  \centering
  \includegraphics[width=6cm]{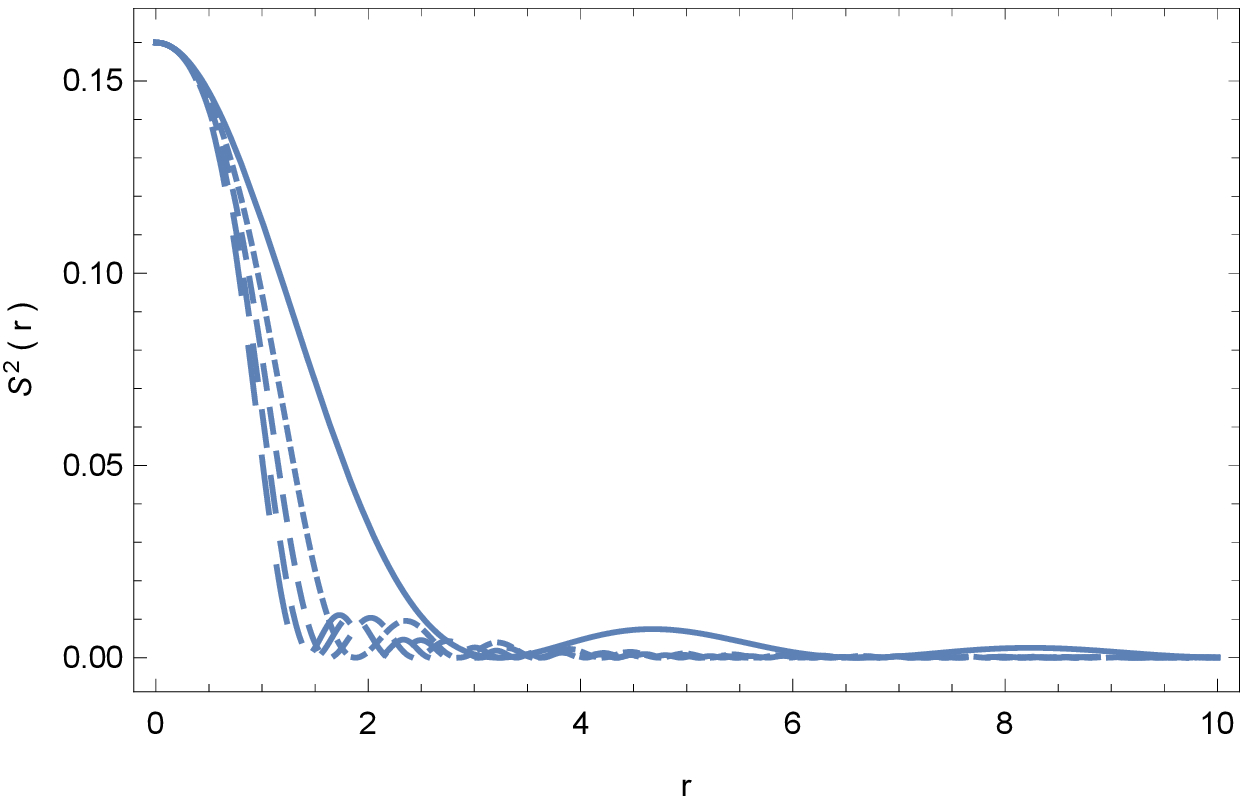}
   \includegraphics[width=6cm]{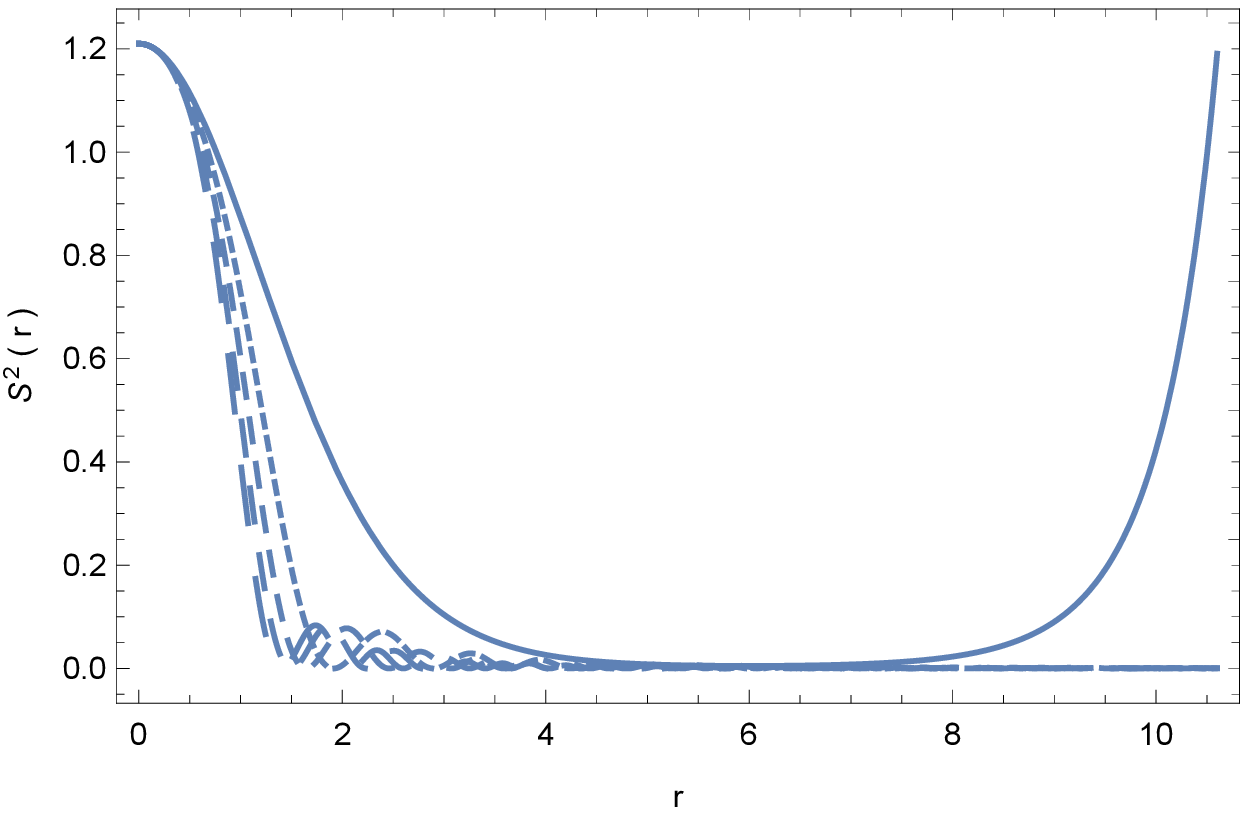}
 \caption{ Variation of the scaled matter energy density $S^2(r)$ as a function of $r$ (in arbitrary units) for $S(0)=0.4$, $V(0)=1.0$ (left panel) and for $S(0)=1.1$, $V(0)=1$ (right panel), and for different values of the dark energy: $\lambda =0$ (solid curve), $\lambda =10$ (dotted curve), $\lambda =20$ (short dashed curve), $\lambda =30$ (dashed curve), and $\lambda =40$ (long dashed curve).}\label{fig2b}
\end{figure*}

  The matter density $\rho $ is defined quantum-mechanically according to  $\rho =m\left|\psi \left(\vec{r}\right)\right|^2=\left(\hbar ^2/8\pi Gm^2\right)S^2\left(\vec{r}\right)$. The variation of $S^2(r)$ as a function of the radial coordinate $r$ is represented, for different values of the cosmological constant $\lambda$, and for two particular set of initial conditions, in Fig.~\ref{fig2b}. We will first discuss the case $\lambda =0$, and $S(0)=0.4$, $V(0)=1.0$, respectively. The matter density has its maximum value at $r=0$, and it decreases rapidly with increasing $r$. However, after reaching a minimum value, the density increases again, attaining a second maximum with a much smaller amplitude, with this behavioral pattern repeating itself up to point where $\rho \approx 0$. We interpret these transitions from a decreasing to an increasing density as corresponding to the existence of a density bounce, and to the oscillations of the density of the quantum matter. This type of bouncing behavior is significantly affected by the presence of the large values of the cosmological constant. The matter density, having its maximum at $r=0$ independently of the absence or presence of the cosmological constant, reaches its first minimum value at much smaller values of $r$, as compared to the $\lambda =0$ case. Moreover, the successive maximums/minimums occur much closer to the origin, and a large number of density bounces do appear, corresponding to matter density oscillations, as compared to the few present in the $\lambda =0$ model. The bouncing behavior essentially depends not only on $\lambda$, but also on the initial conditions, as one can clearly see from the right panel of Fig.~\ref{fig2b}. For the initial values  $S(r)$, $S(0)=1.1$, $V(0)=1$, leading to the blow-up of the wave function, in the case $\lambda =0$ there is a clear density bounce, with the matter density decreasing to a minimum (almost zero) value, and then blowing up for larger values of $r$. This behavior, corresponding to a single bounce, and the appearance of a singularity in the matter density,  is drastically modified by the presence of the cosmological constant that wipes out the singularity in the matter density. Hence, $\rho $ tends to zero through an oscillatory process, with the amplitude of the oscillations slowly decreasing with increasing $r$.

 \subsection{Comparison with the semianalytical solutions}

\begin{figure*}[ht]
  \centering
  \includegraphics[width=6cm]{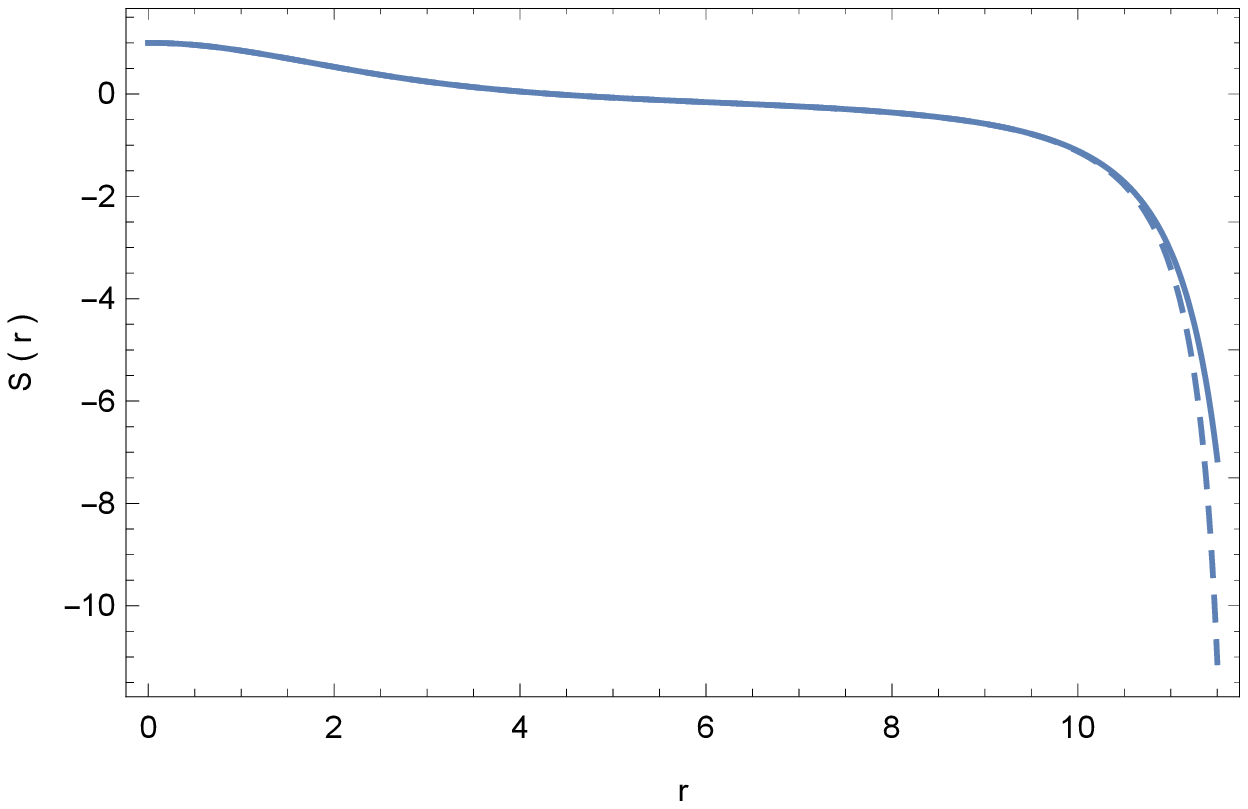}
  \includegraphics[width=6cm]{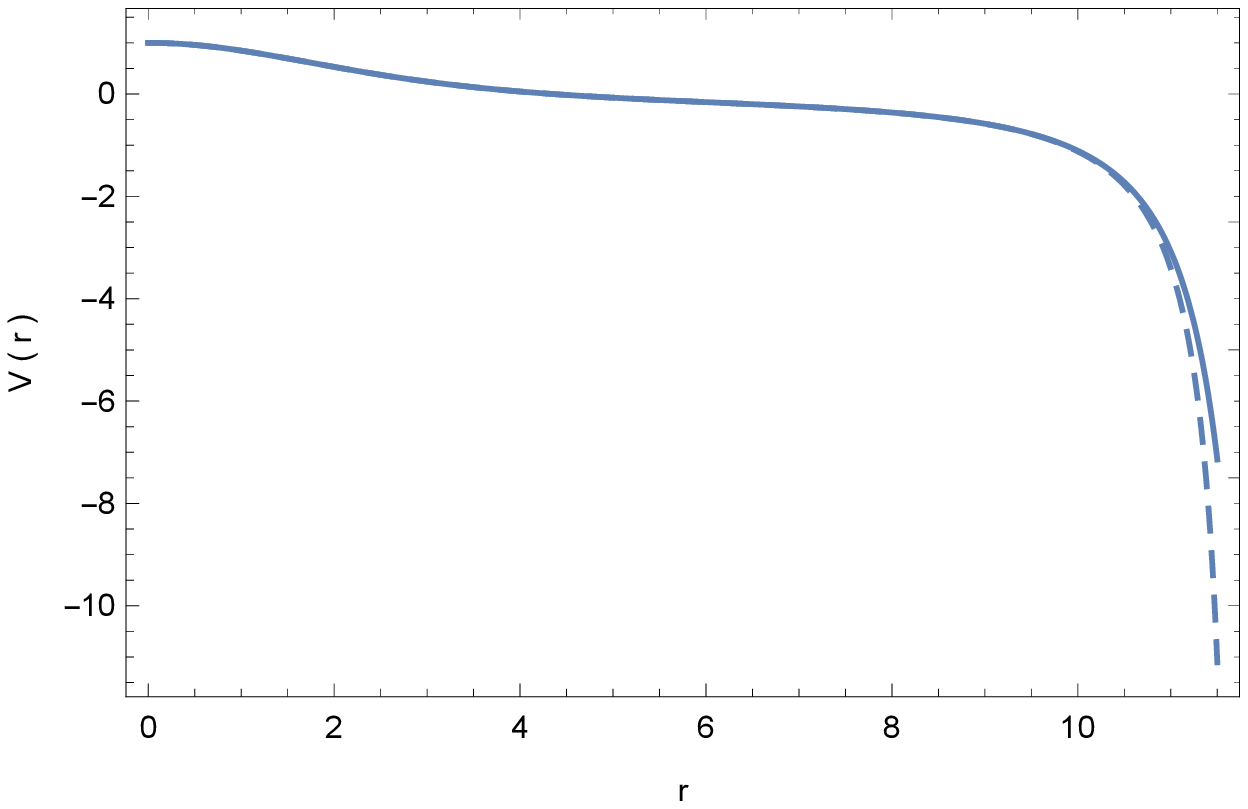}
  \caption{Comparison of the semianalytical solution of the Newton-Schr\"{o}dinger system (dashed curve), with $\lambda =0$, and the full numerical solution (solid curve),  for $S(r)$ (left panel) and $V(r)$ (right panel) (in arbitrary units), for $S(0)=1$ and $V(0)=1$.}\label{fig3}
\end{figure*}

 We will consider now the comparison of the semianalytical solutions of the S-N-$\Lambda$ system, obtained via the Adomian Decomposition Method-Pad\'{e} approximants, and the  full numerical solution. For the case of the S-N system, with $\lambda =0$, the comparison of the two solutions is represented, for $S(0)=1$ and $V(0)=1$, in Fig.~\ref{fig3}.

\begin{figure*}[ht]
  \centering
  \includegraphics[width=6cm]{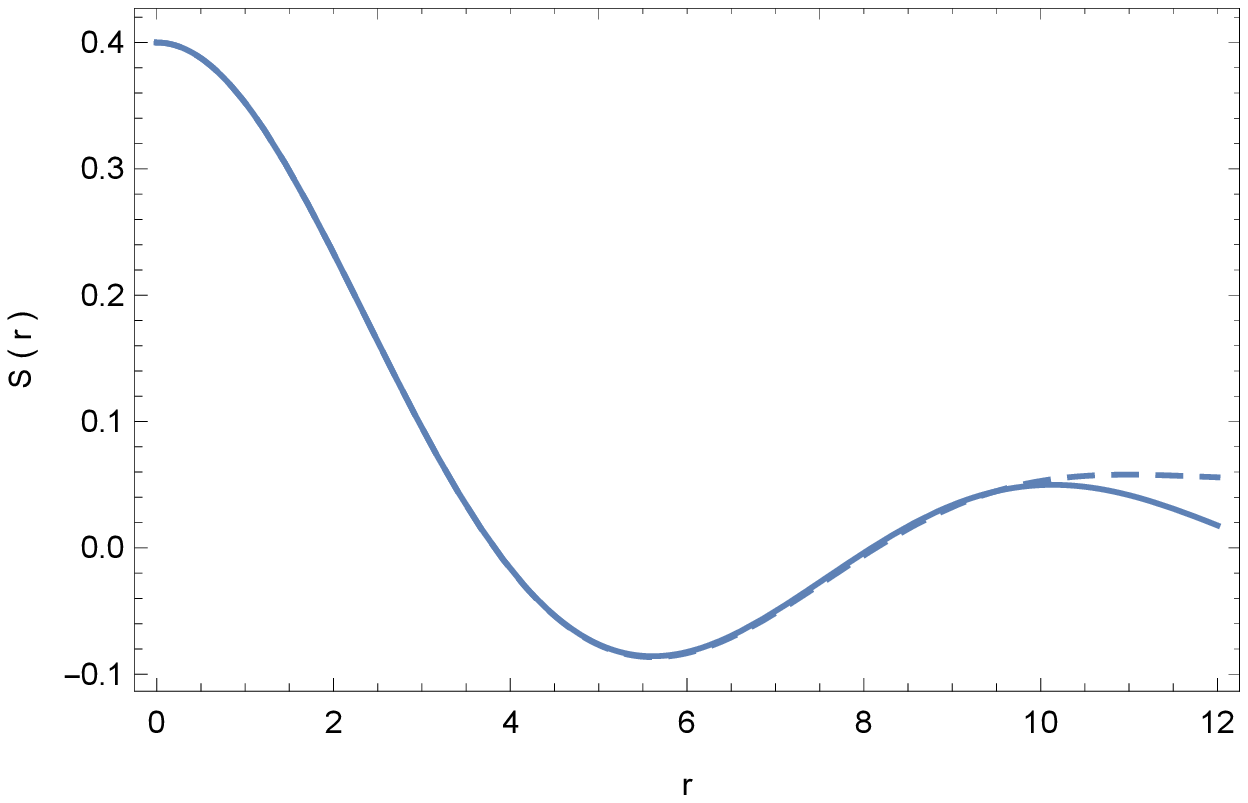}
  \includegraphics[width=6cm]{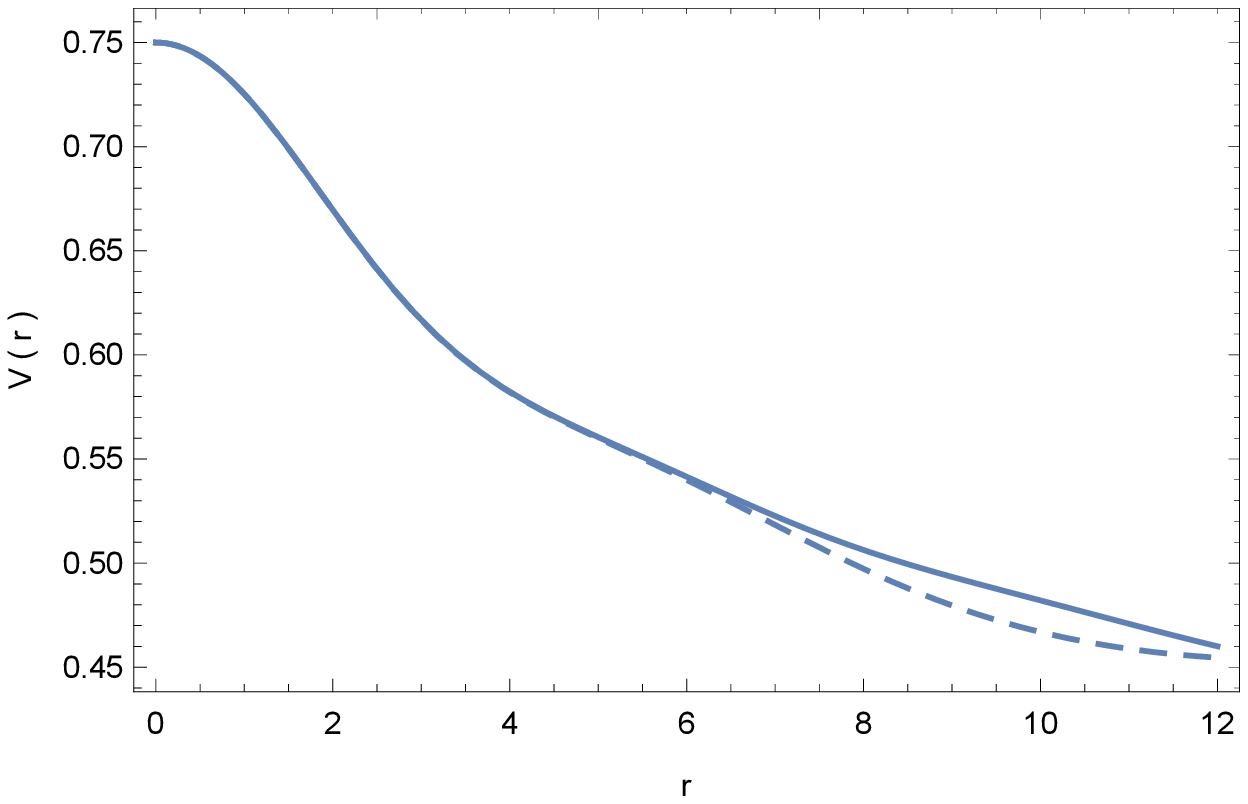}
  \caption{Comparison of the semianalytical solution of the Newton-Schr\"{o}dinger -$\Lambda $ system (dashed curve), and the  full numerical solution (solid curve),  for $S(r)$ (left panel) and $V(r)$ (right panel) (in arbitrary units),  with $\lambda =0.0001$, and $S(0)=0.40$ and $V(0)=0.75$.}\label{fig4}
\end{figure*}

For the adopted initial conditions the semianalytical solution gives an excellent approximation of the numerical up to the appearance of the first singular point. For the case $\lambda =0.0001$, the comparison between the numerical and the semianalytical solution is presented in Fig.~\ref{fig4}, for $S(0)=0.40$ and $V(0)=0.75$.

\begin{figure*}[ht]
  \centering
  \includegraphics[width=6cm]{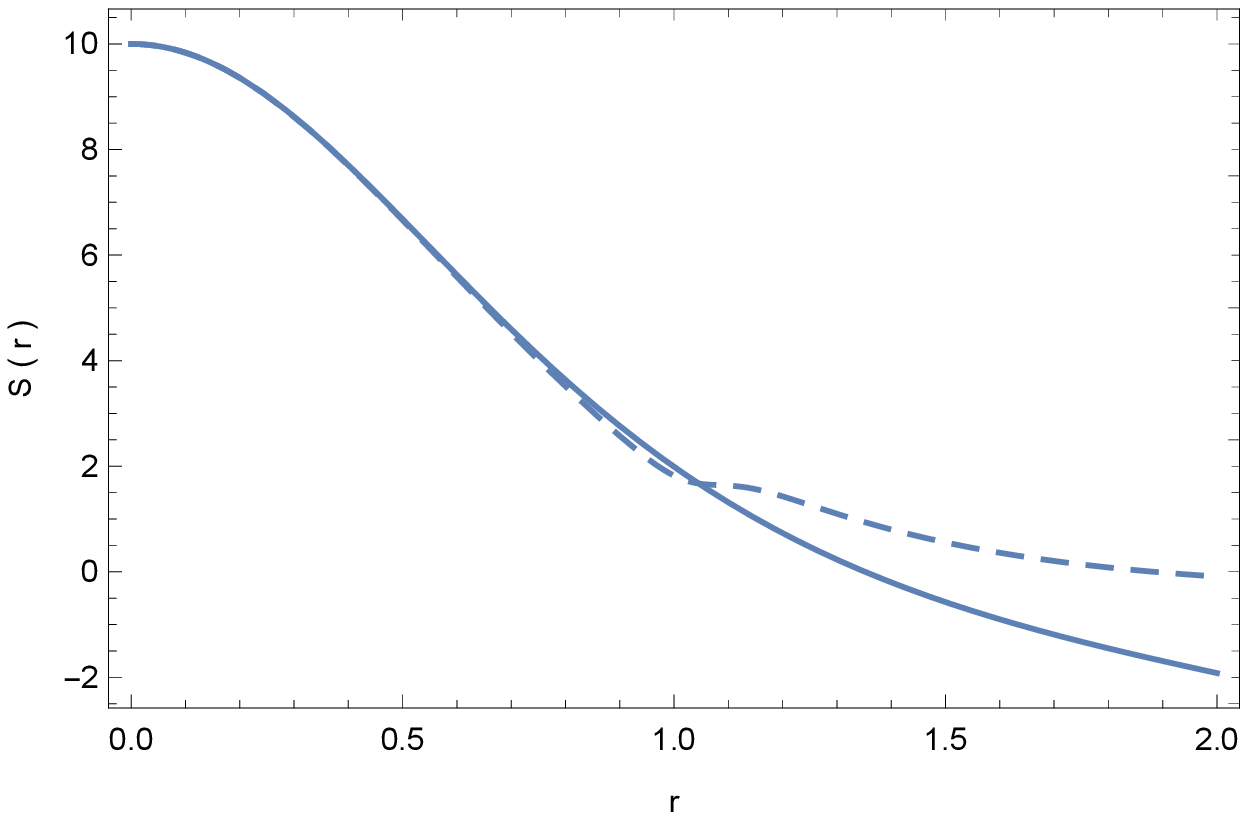}
  \includegraphics[width=6cm]{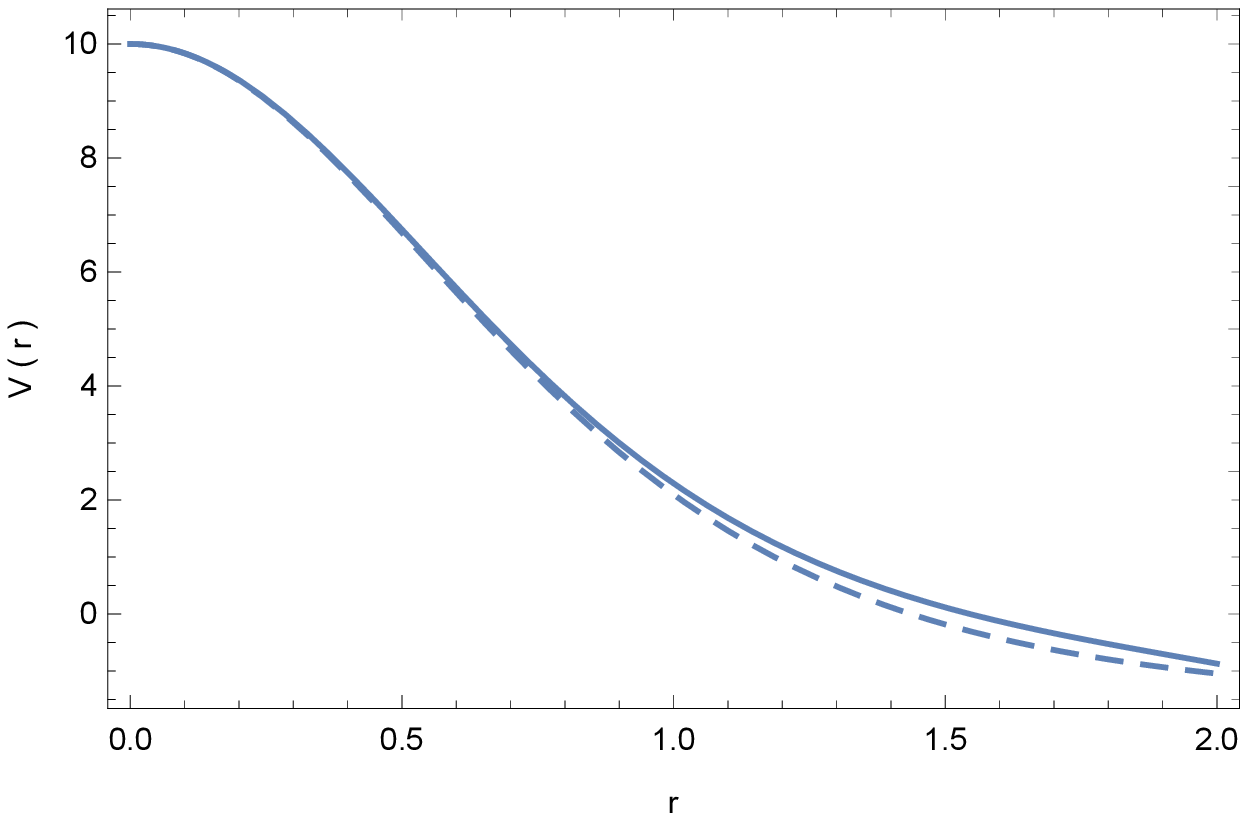}
  \caption{Comparison of the semianalytical solution of the Newton-Schr\"{o}dinger -$\Lambda $ system (dashed curve), and the  full numerical solution (solid curve),  for $S(r)$ (left panel) and $V(r)$ (right panel) (in arbitrary units),  for $\lambda =1.4$, and $S(0)=10$ and $V(0)=10$, respectively.}\label{fig5}
\end{figure*}

 The comparison of the Adomian Decomposition semianalytical solution and the full numerical solution of the Schr\"{o}dinger-Newton-$\Lambda$ system for $\lambda =1.4$ is represented in Fig.~\ref{fig5}.

 For large values of $\lambda$, $S(0)$ and $V(0)$, we obtain a good approximation of the numerical solution in the range $0\leq r\leq 1$.  Even that generally the Adomian series does converge towards the exact/full numerical solution, for large values of $S(0)$, $V(0)$, and $\lambda$, the convergence is slow, and for the few terms considered in the series expansion, the semianalytical solution may describes well the full numerical solution only for small values of $r$, usually in the range $0\leq r\leq 1$. For smaller values of $\lambda$ there is a good concordance between the numerical and the semianalytical approaches, and the two overlap even for large values of $r$. On the other hand it is important to point out that the Adomian Decomposition Method generally fails in the vicinity of the singular points of a differential equation, where the solution diverges.

Hence even in the presence of the cosmological constant, the Adomian Decomposition Methods, used together with the Pad\'{e} approximants approach, provides a good approximation of the  full numerical solution. The precision of the approximation essentially depends on the number of terms included in the Adomian decomposition, as well as of the order $mn$ of the Pad\'{e} approximants. Moreover, the procedure can be easily implemented by using symbolic calculation software that allows to approximate the numerical solution with an arbitrary precision, thus helping in obtaining a deeper insight into its physical properties.

\section{Discussions and final remarks}\label{sect5}

In the present Letter we have considered some basic properties of the simplest extension of the static Schr\"{o}dinger-Newton system in spherical symmetry, which consists in the modification of the Poisson equation through the addition of the dark energy term, which we modeled as a simple cosmological constant. In the present model, the dynamical behavior  of a quantum particle is determined by the nonlinear Schr\"{o}dinger equation containing
an effective potential including the standard Newtonian gravity and the dark energy contributions. After reformulating the S-N-$\Lambda$ system of two differential equations as an equivalent system of integral equations, we have applied the Adomian Decomposition Method to obtain a semianalytic power series solution. In order to avoid the oscillating/singular behaviors in the Adomian series we have approximated them as rational functions via the method of the Pad\'{e} approximants. We have investigated the general S-N-$\Lambda$ system numerically, in order to point out the important role the dark energy may play in the quantum description of gravity, especially if interpreted as a vacuum fluctuation.  The dark energy influences the number of zeros of the wave function, as well as the blow-up radius of the system. But in order to fully describe the effects of the dark energy a complete and systematic numerical and analytical investigation of the system is necessary. In particular, the novel and interesting aspects related to the energy spectrum must be carefully analyzed, and especially in the  limit of  a large cosmological constant value, the spectrum may give some insights into the quantum nature of gravity. For example, the problem of the existence of bound states at infinity takes a different form due to the presence of the nonzero cosmological constant at infinity. Moreover,  the eigenvalue problem for large $r$, from which in principle one can determine exactly the eigenvalues, as well as the eigenfunctions asymptotically in the form of expansions of increasing accuracy becomes very different in the case of the S-N-$\Lambda$ system, as compared to the case of the S-N system. The energy eigenvalues in the presence of the cosmological constant can be obtained from  Eq.~(\ref{act}), by adopting, for example, for the first approximation of the ground state wave function the hydrogen atom wave function, $\psi (r)=\left(1/\pi \sigma ^3\right)^{1/2} e^{-r/\sigma}$ \cite{T3a}, which would allow to obtain perturbatively the energy eigenvalues, and the wave function in the presence of dark energy. But more precise predictions of the model also do depend on a  full numerical study of the S-N-$\Lambda$ system in different physical contexts. The time-dependent S-N-$\Lambda$ system can also be investigated by using the Adomian Decomposition Method, and series solutions of the equation can be obtained easily \cite{Harko}.

 The present results on the existence/nonexistnce of a bounce in the matter density may have interesting cosmological implications. Bouncing solutions in which the Universe smoothly bounce from a collapsing to an expanding phase have attracted a lot of attention recently. A bouncing Universe does appear in Loop Quantum Cosmology, a quantum theory of gravity in which the macroscopic physical quantities (energy density, curvature, etc.), have finite upper bounds \cite{LQC}. Hence,  a contracting homogeneous and isotropic Universe will bounce back to an expanding one at finite values of the scale factor and energy density, thus preventing the appearance of a physical singularity. The blow-up of the solutions of the S-N-$\Lambda$ system for various initial conditions may, at first sight, suggest that Newtonian quantum gravity cannot consistently solve the cosmological problem, and a fine-tuning of the initial conditions is necessary. However, in order to give a full answer to this question the investigation of the cosmological behavior of the time-dependent S-N-$\Lambda$ is necessary, in which the effects of classical general relativity are also included. Such a study  may lead to a better understanding of the quantum gravity aspects in the early Universe, and provide observational signatures of quantum gravity, which may have some observational imprints on the primordial power spectrum of the Cosmic Background Microwave Radiation.

Moreover, the S-N-$\Lambda$ model opens the possibility for the understanding of the quantum to classical transition in the presence of dark energy, which provides a new effect not directly related to the increase of the mass of the particle. And, equally importantly, this quantum model combining classical gravity and quantum mechanics allows the investigation of quantum situations in which not only the gravitational field but also quantum fluctuations (interpreted as a dark energy) play a dominant role.

\section*{Acknowledgments}

We would like to thank the two anonymous referees for comments and suggestions that helped us to improve our manuscript.

\end{document}